\documentclass[aps,pre,twocolumn,showpacs,superscriptaddress]{revtex4}
\usepackage{epsfig}
\usepackage{times}
\begin{document}

\title{Different perceptions of social dilemmas: Evolutionary multigames in structured populations}

\author{Zhen Wang}
\affiliation{School of Software, Dalian University of Technology, Dalian 116621, China}

\author{Attila Szolnoki}
\email{szolnoki.attila@ttk.mta.hu}
\affiliation{Institute of Technical Physics and Materials Science, Research Centre for Natural Sciences, Hungarian Academy of Sciences, P.O. Box 49, H-1525 Budapest, Hungary}

\author{Matja{\v z} Perc}
\email{matjaz.perc@uni-mb.si}
\affiliation{Faculty of Natural Sciences and Mathematics, University of Maribor, Koro{\v s}ka cesta 160, SI-2000 Maribor, Slovenia}
\affiliation{Department of Physics, Faculty of Science, King Abdulaziz University, Jeddah, Saudi Arabia}

\begin{abstract}
Motivated by the fact that the same social dilemma can be perceived differently by different players, we here study evolutionary multigames in structured populations. While the core game is the weak prisoner's dilemma, a fraction of the population adopts either a positive or a negative value of the sucker's payoff, thus playing either the traditional prisoner's dilemma or the snowdrift game. We show that the higher the fraction of the population adopting a different payoff matrix, the more the evolution of cooperation is promoted. The microscopic mechanism responsible for this outcome is unique to structured populations, and it is due to the payoff heterogeneity, which spontaneously introduces strong cooperative leaders that give rise to an asymmetric strategy imitation flow in favor of cooperation. We demonstrate that the reported evolutionary outcomes are robust against variations of the interaction network, and they also remain valid if players are allowed to vary which game they play over time. These results corroborate existing evidence in favor of heterogeneity-enhanced network reciprocity, and they reveal how different perceptions of social dilemmas may contribute to their resolution.
\end{abstract}

\maketitle

\section{Introduction}
Social dilemmas are situations in which individuals are torn between what is best for them and what is best for the society. If selfishness prevails, the pursuit of short-term individual benefits may quickly result in loss of mutually rewarding cooperative behavior and ultimately in the tragedy of the commons \cite{hardin_g_s68}. Evolutionary game theory \cite{maynard_82, weibull_95, hofbauer_98, mestertong_01, nowak_06} is the most commonly adopted theoretical framework for the study of social dilemmas, and none has received as much attention as the prisoner's dilemma game \cite{fudenberg_e86, nowak_n93, santos_prl05, imhof_pnas05, santos_pnas06, tanimoto_pre07, fu_epjb07, gomez-gardenes_prl07, poncela_njp07, fu_pre08b, poncela_epl09, fu_pre09, fu_jtb10, antonioni_pone11, tanimoto_pre12, press_pnas12, hilbe_pnas13, szolnoki_pre14}. Each instance of the game is contested by two players who have to decide simultaneously whether they want to cooperate or defect. The dilemma is given by the fact that although mutual cooperation yields the highest collective payoff, a defector will do better if the opponent decides to cooperate.

Since widespread cooperation in nature is one of the most important challenges to Darwin's theory of evolution and natural selection, ample research has been devoted to the identification of mechanisms that may lead to a cooperative resolution of social dilemmas. Classic examples reviewed in \cite{nowak_s06} include kin selection \cite{hamilton_wd_jtb64a}, direct and indirect reciprocity \cite{trivers_qrb71, axelrod_s81}, network reciprocity \cite{nowak_n92b}, as well as group selection \cite{wilson_ds_an77}. Recently, however, interdisciplinary research linking together knowledge from biology and sociology as well as mathematics and physics has revealed many refinements to these mechanisms and also new ways by means of which the successful evolution of cooperation amongst selfish and unrelated individuals can be understood \cite{szabo_pr07, roca_plr09, schuster_jbp08, perc_bs10, santos_jtb12, perc_jrsi13, rand_tcs13}.

One of the more recent and very promising developments in evolutionary game theory is the introduction of so-called multigames \cite{hashimoto_jtb06, hashimoto_jtb14} or mixed games \cite{wardil_csf13} (for earlier conceptually related work see \cite{cressman_igtr00}), where different players in the population adopt different payoff matrices. Indeed, it is often the case that a particular dilemma is perceived differently by different players, and this is properly taken into account by considering a multigame environment. A simple example to illustrate the point entails two drivers meeting in a narrow street and needing to avoid collision. However, while the first driver drives a cheap old car, the second driver drives a brand new expensive car. Obviously, the second driver will be more keen on averting a collision. Several other examples could be given to illustrate that, when we face a conflict, we are likely to perceive differently what we might loose in case the other player chooses to defect. The key question then is, how the presence of different payoff matrices, motivated by the different perception of a dilemma situation, will influence the cooperation level in the whole population?

Multigames were thus far studied in well-mixed systems, but since stable solutions in structured populations can differ significantly -- a prominent example of this fact being the successful evolution of cooperation in the prisoner's dilemma game through network reciprocity \cite{nowak_n92b} -- it is of interest to study multigames also within this more realistic setup. Indeed, interactions among players are frequently not random and best described by a well-mixed model, but rather they are limited to a set of other players in the population and as such are best described by a network \cite{doebeli_el05, szabo_pr07, roca_plr09, perc_bs10, perc_jrsi13}. With this as motivation, we here study evolutionary multigames on the square lattice and scale-free networks, where the core game is the weak prisoner's dilemma while at the same time some fraction of players adopts either a positive or a negative value of the sucker's payoff. Effectively, we thus have some players using the weak prisoner's dilemma payoff matrix, some using the traditional prisoner's dilemma payoff matrix, and also some using the snowdrift game payoff matrix. Within this multigame environment, we will show that the higher the heterogeneity of the population in terms of the adopted payoff matrices, the more the evolution of cooperation is promoted. Furthermore, we will elaborate on the responsible microscopic mechanisms, and we will also test the robustness of our observations. Taken together, we will provide firm evidence in support of heterogeneity-enhanced network reciprocity and show how different perceptions of social dilemmas contribute to their resolution. First, however, we proceed with presenting the details of the mathematical model.

\section{Evolutionary multigames}
We study evolutionary multigames on the square lattice and the Barab\'asi-Albert scale-free network \cite{barabasi_s99}, each with an average degree $k=4$ and size $N$.
These graphs, being homogeneous and strongly heterogeneous, represent two extremes of possible interaction topology.
Each player is initially designated either as cooperator ($C$) or defector ($D$) with equal probability. Moreover, each instance of the game involves a pairwise interaction where mutual cooperation yields the reward $R$, mutual defection leads to punishment $P$, and the mixed choice gives the cooperator the sucker's payoff $S$ and the defector the temptation $T$. The core game is the weak prisoner's dilemma, such that $T>1$, $R=1$ and $P=S=0$. A fraction $\rho$ of the population, however, uses different $S$ values to take into account the different perception of the same social dilemma. In particular, one half of the randomly chosen $\rho N$ players uses $S=+\Delta$, while the other half uses $S=-\Delta$, where $0 < \Delta < 1$. We adopt the equal division of positive and negative $S$ values to ensure that the average over all payoff matrices returns the core weak prisoner's dilemma, which is convenient for comparisons with the baseline case. Primarily, we consider multigames where, once assigned, players do not change their payoff matrices, but we also verify the robustness of our results by considering multigames with time-varying matrices.

We simulate the evolutionary process in accordance with the standard Monte Carlo simulation procedure comprising the following elementary steps. First, according to the random sequential update protocol, a randomly selected player $x$ acquires its payoff $\Pi_x$ by playing the game with all its neighbors. Next, player $x$ randomly chooses one neighbor $y$, who then also acquires its payoff $\Pi_y$ in the same way as previously player $x$. Importantly, at each instance of the game the applied payoff matrix is that of the randomly chosen player who collects the payoffs, which may result in an asymmetric payoff allocation depending on who is central. This fact, however, is key to the main assumption that different players perceive the same situation differently. Once both players acquire their payoffs, then player $x$ adopts the strategy $s_y$ from player $y$ with a probability determined by the Fermi function
\begin{equation}
\label{eq1}
W(s_y \to s_x)=\frac{1}{1+\exp[(\Pi_x-\Pi_y)/K]},
\end{equation}
where $K=0.1$ quantifies the uncertainty related to the strategy adoption process \cite{blume_l_geb93, szabo_pr07}. In agreement with previous works, the selected value ensures that strategies of better-performing players are readily adopted by their neighbors, although adopting the strategy of a player that performs worse is also possible \cite{perc_pre08b, szolnoki_njp08}. This accounts for imperfect information, errors in the evaluation of the opponent, and similar unpredictable factors.

Each full Monte Carlo step (MCS) consists of $N$ elementary steps described above, which are repeated consecutively, thus giving a chance to every player to change its strategy once on average. All simulation results are obtained on networks
typically with $N=10^4$ players, but larger system size is necessary on
the proximity to phase transition points, and the fraction of cooperators $f_C$ is determined in the stationary state after a sufficiently long relaxation lasting up to $2 \cdot 10^5$ MCS. To further improve accuracy, the final results are averaged over $200$
independent realizations, including the generation of the scale-free networks, at each set of parameter values.

\section{Results}
Before turning to the main results obtained in structured populations, we first briefly summarize the evolutionary outcomes in well-mixed populations. Although the subpopulation adopting the $T>1$, $R=1$, $P=0$ and $S=+\Delta$ parametrization fulfills $T>R>S>P$, and thus in principle plays the snowdrift game where the equilibrium is a mixed $C+D$ phase, cooperators in the studied multigame actually never survive. Since there are also players who adopt either the weak ($T>R>P=S$) or the traditional ($T>R>P>S$) prisoner's dilemma payoff matrix, the asymmetry in the interactions renders cooperation evolutionary unstable. In fact, in well-mixed populations the baseline case given by the average over all payoff matrices is recovered, which in our setup is the weak prisoner's dilemma, where for all $T>1$ cooperators are unable to survive. More precisely, cooperators using $S=-\Delta$ die out first, followed by those using $S=0$ and $S=+\Delta$, and this ranking is preserved even if the subpopulation using $S=0$ is initially significantly larger than the other two subpopulations (at small $\rho$ values). Although in finite well-mixed populations the rank of this extinction pattern could be very tight, it does not change the final fate of the population to arrive at complete defection.

\begin{figure}
\centerline{\epsfig{file=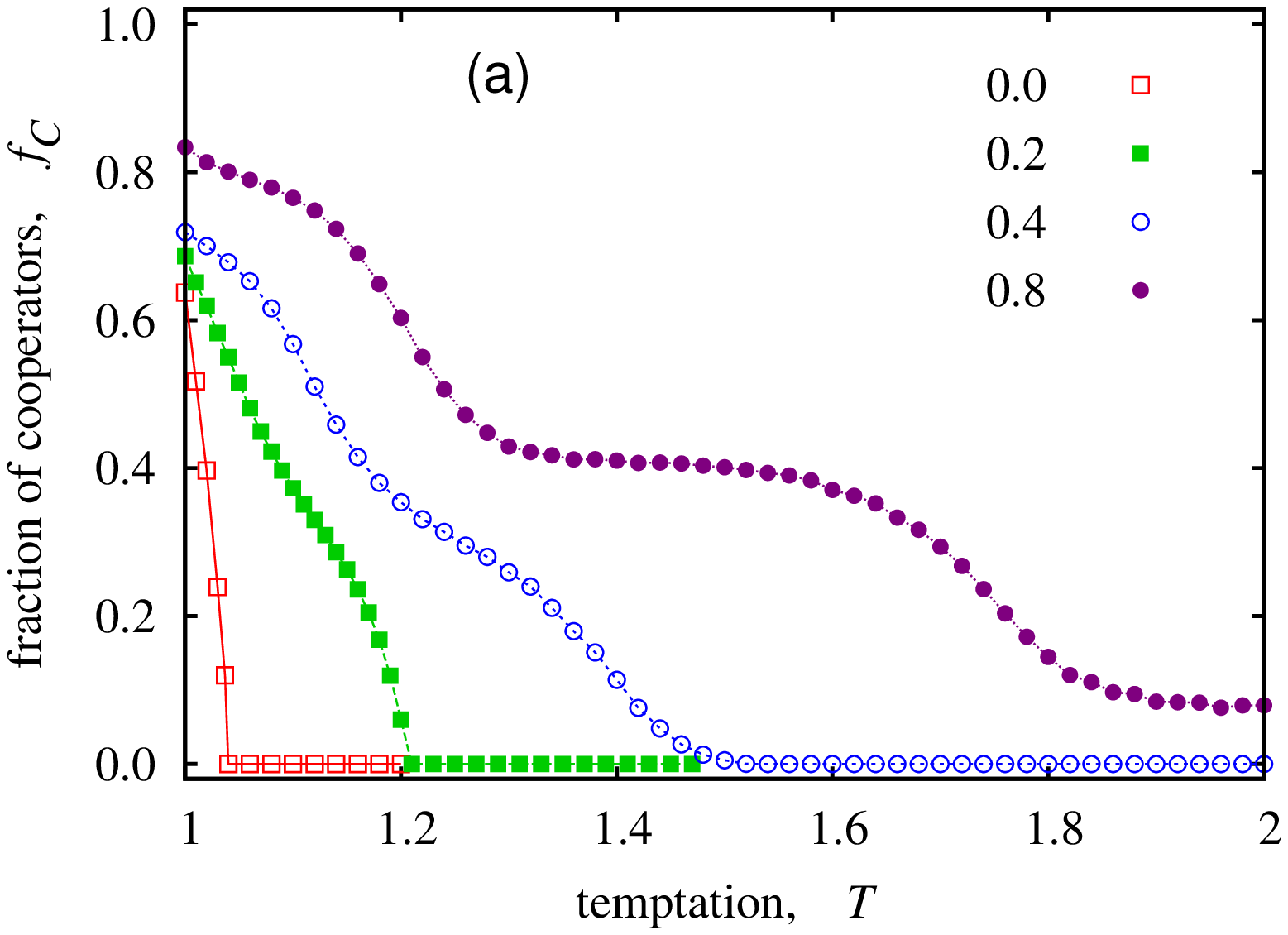,width=8cm}}
\centerline{\epsfig{file=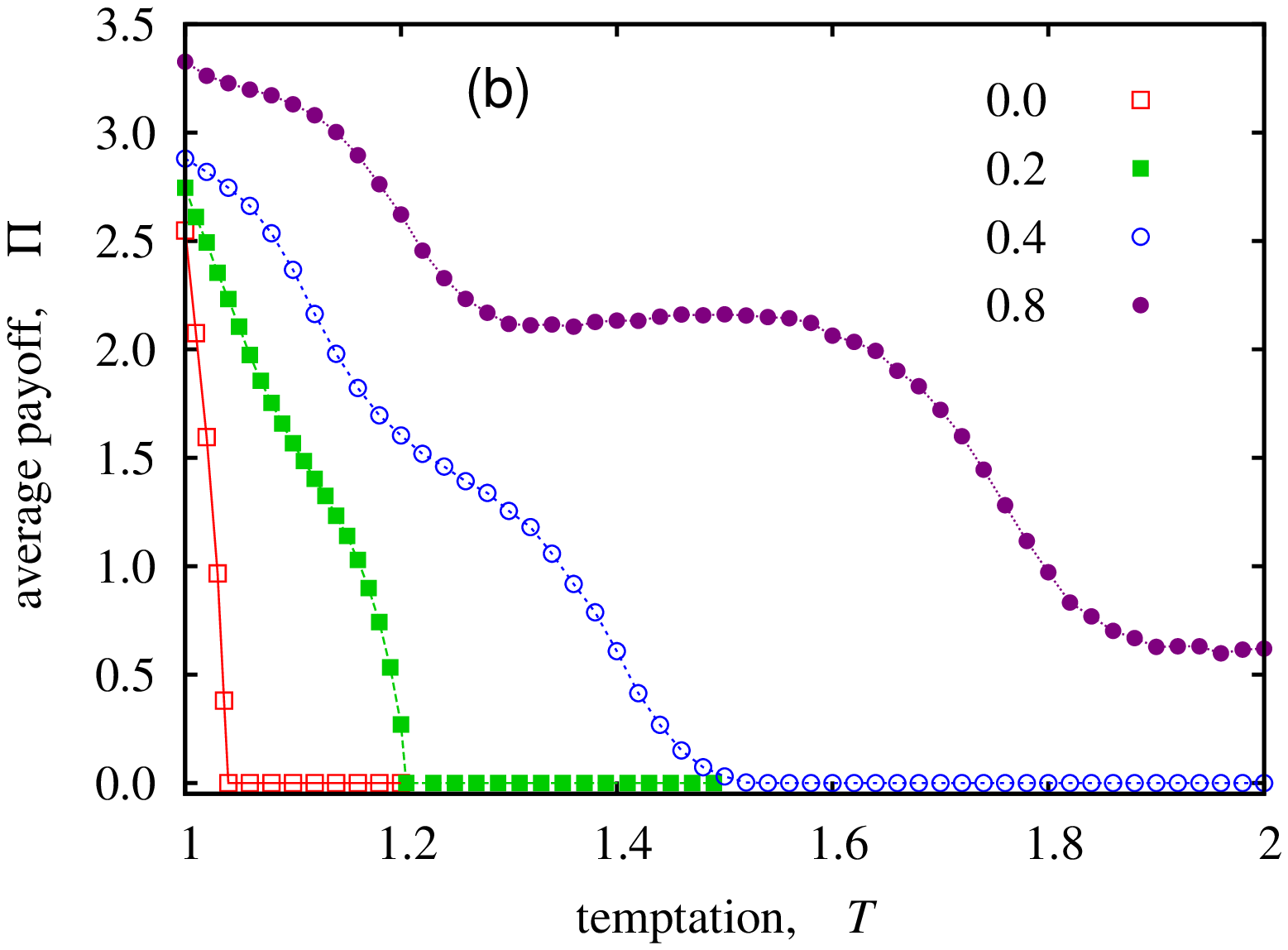,width=8cm}}
\caption{(Color online) Evolution of cooperation (top panel) and the average payoff of the population (bottom panel) in the multigame environment on the square lattice. Depicted results were obtained for $\rho=1$ and different values of $\Delta$, as indicated in the legend. Here $\rho=1$ means that all players use either $S=+\Delta$ or $S=-\Delta$ (none use $S=0$). Larger values of $\Delta$ allow cooperators to survive at larger values of $T$. Importantly, this improvement in $f_C$ is also accompanied by a suitable increase in the average payoff of the population, as shown in the bottom panel.}
\label{Delta}
\end{figure}

In structured populations, as expected from previous experience
\cite{doebeli_el05, szabo_pr07, roca_plr09, perc_bs10, perc_jrsi13}, we can observe different solutions, where cooperators can coexist with defectors over a wide range of parameter values. But more importantly, the multigame environment, depending on $\rho$ and $\Delta$, can elevate the stationary cooperation level significantly beyond that warranted by network reciprocity alone. We first demonstrate this in Fig.~\ref{Delta}(a), where we plot the fraction of cooperators $f_C$ as a function of the temptation value $T$, as obtained for $\rho=1$ and by using different values of $\Delta$. It can be observed that the larger the value of $\Delta$ the larger the value of $T$ at which cooperators are still able to survive. Indeed, for $\Delta=0.8$ cooperation prevails across the whole interval of $T$. Since some players use a negative value of $S$, it is nevertheless of interest to test whether the elevated level of cooperation actually translates to a larger average payoff of the population. It is namely known that certain mechanisms aimed at promoting cooperative behavior, like for example punishment \cite{sigmund_tee07}, elevate the level of cooperation but at the same time fail to raise the average payoff accordingly due to the entailed negative payoff elements. As illustrated in Fig.~\ref{Delta}(b), however, this is not the case at present since larger values of $f_C$ readily translate to larger average payoffs of the population.

\begin{figure}[b]
\centerline{\epsfig{file=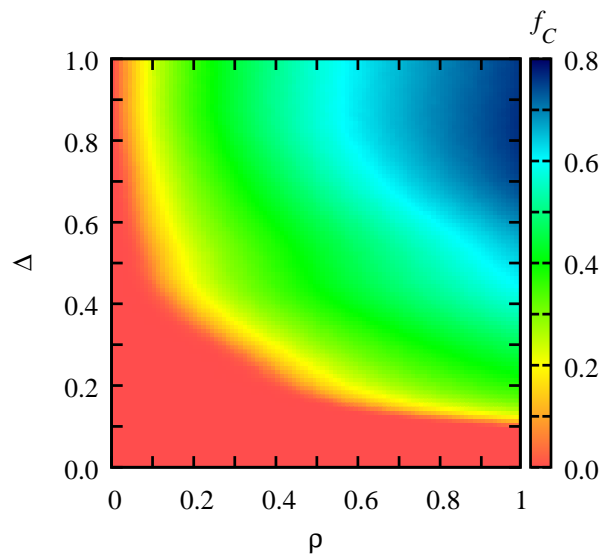,width=8cm}}
\caption{(Color online) Evolution of cooperation in the multigame environment on the square lattice, as obtained in dependence on $\rho$ and $\Delta$. The color map encodes the stationary fraction of cooperators $f_C$. It can be observed that the dependence of $f_C$ on both $\rho$ and $\Delta$ is monotonous, and that it is thus beneficial for the population to be in the most heterogeneous state possible. Depicted results were obtained for $T=1.1$, but qualitatively equal evolutionary outcomes can be observers also for other values of $T$.}
\label{rho}
\end{figure}

In the light of these results, we focus solely on the fraction of cooperators and show in Fig.~\ref{rho} how $f_C$ varies in dependence on $\rho$ and $\Delta$ at a given temptation value  $T$. Presented results indicate that what we have observed in Fig.~\ref{Delta}(a), namely the larger the value of $\Delta$ the better, actually holds irrespective of the value of $\rho$. More to the point, larger $\rho$ values support cooperation stronger, which corroborates the argument that the more heterogeneous the multigame environment the better. Results presented in Fig.~\ref{rho} also suggest that it is better to have many players using higher $S$ values, regardless of the fact that the price is an equal number of players in the population using equally high but negative $S$ values. These observations hold irrespective of the temptation $T$, and they fit well to the established notion that heterogeneity, regardless of its origin, promotes cooperation by enhancing network reciprocity \cite{perc_pre08, santos_n08, szolnoki_epjb08, lei_c_pa10, santos_jtb12, sun_l_ijmpc13, tanimoto_13, vukov_njp12, zhu_p_pone14, maciejewski_pcbi14, yuan_wj_pone14}.

\begin{figure}
\centerline{\epsfig{file=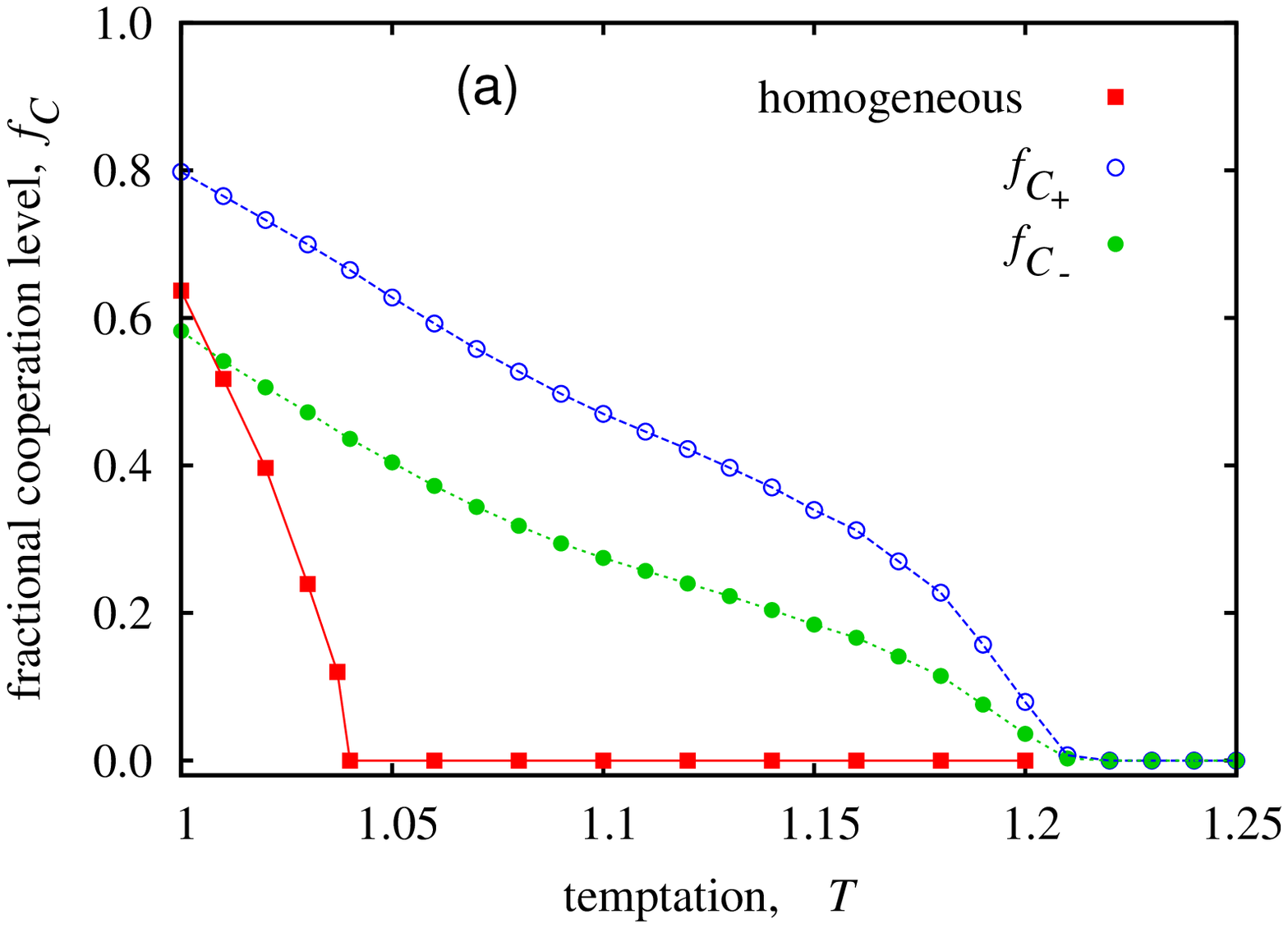,width=8.5cm}}
\centerline{\epsfig{file=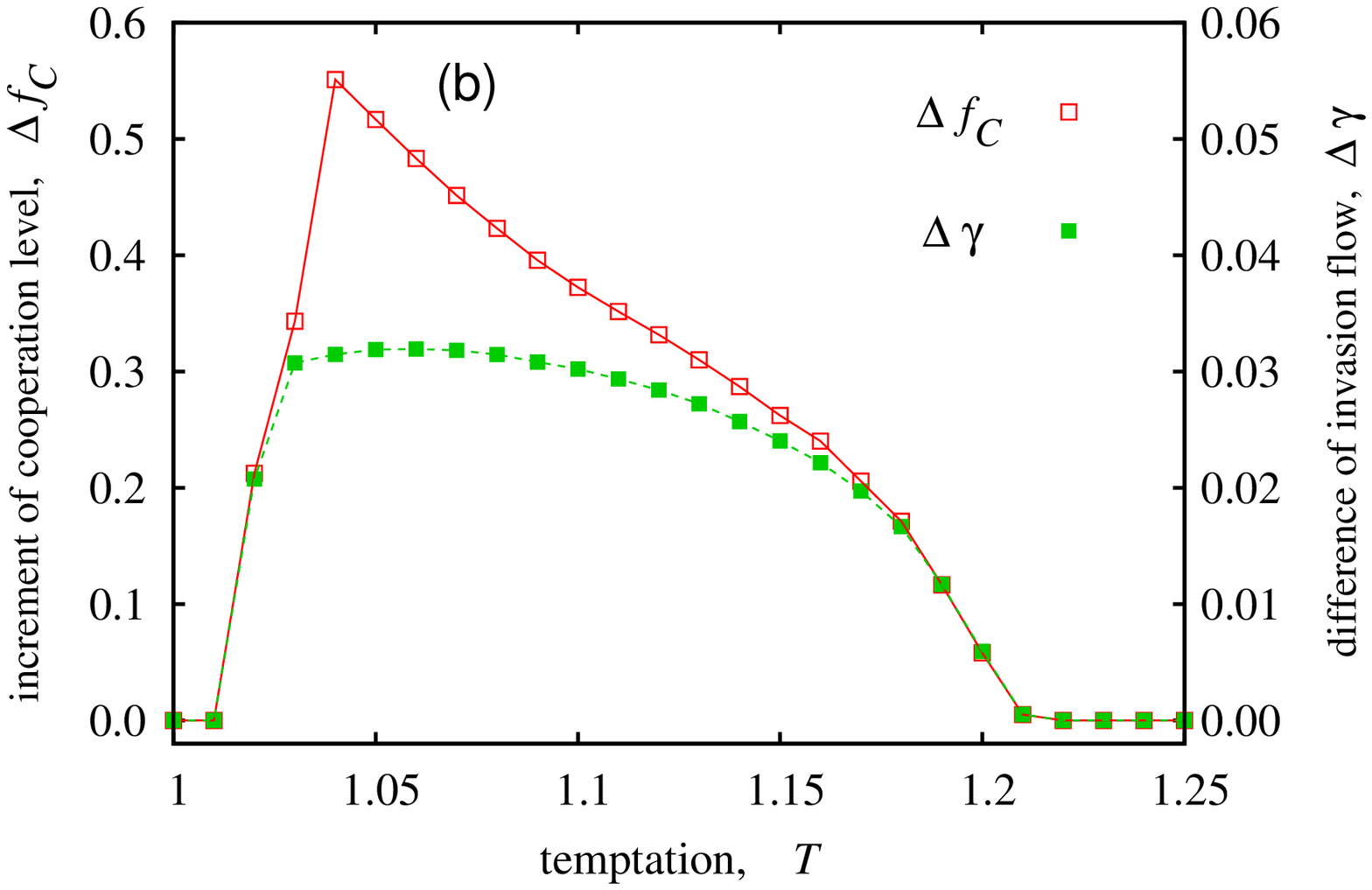,width=8.5cm}}
\caption{(Color online) Top panel depicts the level of cooperation in the two subpopulations on the square lattice, where $f_{C_+}$ denotes the fraction of cooperators among those who use $S=+\Delta$, while $f_{C_-}$ denotes the fraction of cooperators in the group where $S=-\Delta$ is used. For reference, we also plot the cooperation level in the corresponding homogeneous population, where every player uses $S=0$. Expectedly, the level of cooperation is largest in the subpopulation where players use $S=+\Delta$. Much more surprisingly, however, the level of cooperation in the subpopulation where players use $S=-\Delta$ still significantly exceeds the baseline outcome of the homogeneous weak prisoner's dilemma game. Bottom panel depicts the difference between the level of cooperation in the homogeneous and the heterogeneous multigame environment $\Delta f_C$, along with the difference in the strategy invasion flow $\Delta \gamma$ between the two ``+'' and ``-'' subpopulations (see main text for details). These results were obtained with $\rho=1$ and $\Delta=0.2$, but remain qualitatively identical also for other parameter values.}
\label{asymmetry}
\end{figure}

To support these arguments and to pinpoint the microscopic mechanism that is responsible for the promotion of cooperation in the multigame environment, we first monitor the fraction of cooperators within subgroups of players that use different payoff matrices. For clarity, we use $\rho=1$, where only two subpopulations exist (players use either $S=+\Delta$ or $S=-\Delta$, but nobody uses $S=0$), and where the positive effect on the evolution of cooperation is the strongest (see Fig.~\ref{rho}). Accordingly, one group is formed by players who use $S=+\Delta$, and the other is formed by players who use $S=-\Delta$. We denote the fraction of cooperators in these two subpopulations by $f_{C_+}$ and $f_{C_-}$, respectively. As Fig.~\ref{asymmetry}(a) shows, even if only a moderate $\Delta$ value is applied, the cooperation level among players who use a positive $S$ value is significantly higher than among those who use a negative $S$ value. Unexpectedly, even among those players who effectively play a traditional prisoner's dilemma ($T>R>P>S$), the level of cooperation is still much higher than the level of cooperation that is supported solely by network reciprocity (without multigame heterogeneity) in the weak prisoner's dilemma ($T>R>P=S$). This fact further supports the conclusion that the introduction of heterogeneity through the multigame environment involves the emergence of strong cooperative leaders, which further aid and invigorate traditional network reciprocity. Unlike defectors, cooperators benefit from a positive feedback effect, which originates in the subpopulation that uses positive $S$ values and then spreads towards the subpopulation that uses negative $S$ values, ultimately giving rise to an overall higher social welfare (see Fig.~\ref{Delta}(b)).

This explanation can be verified directly by monitoring the information exchange between the two subpopulations. More precisely, we measure the frequency of strategy imitations between players belonging to the two different subpopulations. The difference $\Delta \gamma$ is positive when players belonging to the ``-'' subpopulation adopt the strategy from players belonging to the ``+'' subpopulation more frequently than vice versa. Results presented in Fig.~\ref{asymmetry}(b) demonstrate clearly that the level of cooperation is increased only if there is significant asymmetry in the strategy imitation flow in favor of the ``+'' subpopulation. Such symmetry breaking, which is due to the multigame environment, supports a level of cooperation in the homogeneous weak prisoner's dilemma that notably exceeds the level of cooperation that is supported solely by traditional network reciprocity.

\begin{figure}
\centerline{\epsfig{file=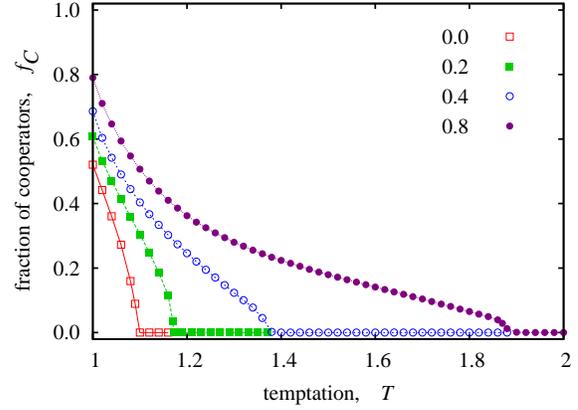,width=8cm}}
\caption{(Color online) Evolution of cooperation in the multigame environment on the scale-free network with degree-normalized payoffs. Depicted results were obtained when only 2\% of the hubs (high-degree players) used $S=+\Delta$, while the rest of the population used a moderately negative $S$ value (see main text for details). As on the square lattice (see top panel of Fig.~\ref{Delta}), larger values of $\Delta$ (see legend) allow cooperators to survive at larger values of $T$.}
\label{sf}
\end{figure}

We proceed by testing the robustness of our observations and expanding this study to heterogeneous interaction networks. First, we consider the Barab{\'a}si-Albert scale-free network \cite{barabasi_s99}, where influential players are a priori present due to the heterogeneity of the topology. Previous research, however, has shown that the positive impact of degree heterogeneity vanishes if payoffs are normalized with the degree of players, as to account for the elevated costs of participating in many games \cite{santos_jeb06, masuda_prsb07, tomassini_ijmpc07, szolnoki_pa08}. We therefore apply degree-normalized payoffs to do away with cooperation promotion that would be due solely to the heterogeneity of the topology. Furthermore, by striving to keep the average over all payoff matrices equal to the weak prisoner's dilemma, it is important to note that the heterogeneous interaction topology allows us to introduce only a few strongly connected players into the $S=+\Delta$ subpopulation, while the rest can use only a moderately negative $S$ value. Specifically, we assigned $S_1=+\Delta$ to only 2\% of the hubs, while the rest used $S_2=-0.0204 \cdot S_1$ to fulfill $0.02 \cdot S_1+0.98 \cdot S_2=0$ (average over all $S$ in the population  equal to zero to yield, on average, the weak prisoner's dilemma payoff ranking). As results depicted in Fig.~\ref{sf} show, even with this relatively minor modification that introduces the multigame environment, the promotion of cooperation is significant if only $\Delta$ is sufficiently large (see legend). Evidently, $\Delta=0$ returns the modest cooperation level that has been reported before on scale-free networks with degree-normalized payoffs, but for $\Delta=0.8$ the coexistence of cooperators and defectors is possible almost across the whole interval of $T$. It is also important to note that the positive effect could be easily amplified  further simply by introducing more players into the $S=+\Delta$ subpopulation and letting the remainder use an accordingly even less negative values of $S$. These results indicate that the topology of the interaction network has only secondary importance, because the heterogeneity that is introduced by payoff differences already provides the necessary support for the successful evolution of cooperation. Consequently, in the realm of the introduced multigame environment, we have observed qualitatively identical cooperation-supporting effects when using the random regular graph or the configurational model of Bender and Canfield \cite{bender, bollobas, molloy} for generating the interaction network.

\begin{figure}
\centerline{\epsfig{file=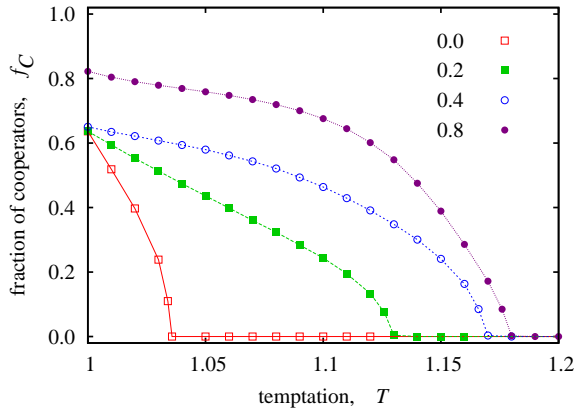,width=8cm}}
\caption{(Color online) Evolution of cooperation in the time-varying multigame environment on the square lattice. Depicted results were obtained when players could choose between $S=+\Delta$ and $S=-\Delta$ with equal probability at each instance of the game (see legend for the applied $\Delta$ values). As on the square lattice with time invariable subpopulations (see top panel of Fig.~\ref{Delta}), larger values of $\Delta$ allow cooperators to survive at larger values of $T$, although in this case the positive impact on the evolution of cooperation is less strong.}
\label{temporary}
\end{figure}

Lastly, we present results obtained within a time-varying multigame environment to further corroborate the robustness of our main arguments. Several examples could be provided as to why players' perception might change over time. The key point is that players may still perceive the same dilemma situation differently, and hence they may use different payoff matrices. Our primary goal here is to present the results obtained with a minimal model, although extensions towards more sophisticated and realistic models are of course possible. Accordingly, unlike considered thus far, players do not have a permanently assigned $S$ value, but rather, they can choose between $S=+\Delta$ and $S=-\Delta$ with equal probability at each instance of the game. Naturally, this again returns the $S=0$ weak prisoner's dilemma on average over time, and as shown in \cite{wardil_csf13}, in well-mixed populations returns the complete defection stationary state. In structured populations, however, for $\Delta>0$, we can again observe promotion of cooperation beyond the level that is warranted solely by network reciprocity. For simplicity, results presented in Fig.~\ref{temporary} were obtained by using the square lattice as the underlying interaction network, but in agreement with the results presented in Fig.~\ref{sf}, qualitatively identical evolutionary outcomes are obtained also on heterogeneous interaction networks. Comparing to the results presented in Fig.~\ref{Delta}(a), where the time invariable multigame environment was applied, we conclude that in the time-varying multigame environment the promotion of cooperation is less strong. This, however, is understandable, since the cooperation-supporting influential players emerge only for a short period of time, but on average the overall positive effect in the stationary state is still clearly there. To conclude, it is worth pointing out that time-dependent perceptions of social dilemmas open the path towards coevolutionary models, as studied previously in the realm of evolutionary games \cite{zimmermann_pre04, szolnoki_njp09, ohtsuki_prl07, perc_bs10, cardillo_njp10, moyano_jtb09}, and they also invite the consideration of the importance of time scales \cite{roca_prl06} in evolutionary multigames.

\section{Discussion}
We have studied multigames in structured populations under the assumption that the same social dilemma is often perceived differently by competing players, and that thus they may use different payoff matrices when interacting with their opponents. This essentially introduces heterogeneity to the evolutionary game and aids network reciprocity in sustaining cooperative behavior even under adverse conditions. As the core game and the baseline for comparisons, we have considered the weak prisoner's dilemma, while the multigame environment has been introduced by assigning to a fraction of the population either a positive or a negative value of the sucker's payoff. We have shown that, regardless of the structure of the interaction network, and also irrespective of whether the multigame environment is time invariant or not, the evolution of cooperation is promoted the more the larger the heterogeneity in the population. As the responsible microscopic mechanism behind the enhanced level of cooperation, we have identified an asymmetric strategy imitation flow from the subpopulation adopting positive sucker's payoffs to the population adopting negative sucker's payoffs. Since the subpopulation where players use positive sucker's payoffs expectedly features a higher level of cooperation, the asymmetric strategy imitation flow thus acts in favor of cooperative behavior also in the other subpopulations, and ultimately it raises the overall level of social welfare in the population.

The obtained results in structured populations are in contrast to the results obtained in well-mixed populations, where simply the baseline weak prisoner's dilemma is recovered regardless of multigame parametrization. Although it is expected that structured populations support evolutionary outcomes that are different from the mean-field case \cite{szabo_pr07, roca_plr09, perc_bs10, perc_jrsi13}, the importance of this fact for multigames is of particular relevance since interactions among players are frequently not best described by a well-mixed model, but rather they are limited to a set of other players in the population and as such are best described by a network. Put differently, although sometimes analytically solvable, the well-mixed models can at best support proof-of-principle studies, but otherwise have limited applicability for realistic systems.

Taken together, the presented results add to the existing evidence in favor of heterogeneity-enhanced network reciprocity, and they further establish heterogeneity among players as a strong fundamental feature that can elevate the cooperation level in structured populations past the boundaries that are imposed by traditional network reciprocity. The rather surprising role of different perceptions of the same conflict thus reveals itself as a powerful mechanism for resolving social dilemmas, although it is rooted in the same fundamental principles as other mechanisms for cooperation promotion that rely on heterogeneity. We hope this paper will motivate further research on multigames in structured populations, which appears to be an underexplored subject with many relevant implications.

\begin{acknowledgments}
This research was supported by the Hungarian National Research Fund (Grant K-101490), TAMOP-4.2.2.A-11/1/KONV-2012-0051, the Slovenian Research Agency (Grants J1-4055 and P5-0027), and the Fundamental Research Funds for Central Universities (Grant DUT13LK38).
\end{acknowledgments}

\end{document}